\documentclass[showpacs]{revtex4}
\usepackage{graphicx}
\begin{document}
\title{A consistent quasiparticle picture of Quark-Gluon plasma
and the velocity of sound}
\author{Sanjay K. Ghosh}
\email{sanjay@bosemain.boseinst.ac.in}
\author{Tamal K. Mukherjee}
\email{tamal@bosemain.boseinst.ac.in}
\author{Sibaji Raha}
\email{sibaji@bosemain.boseinst.ac.in}
\affiliation{Department of Physics, Bose Institute, 93/1, A.P.C Road, 
Kolkata - 700 009, INDIA}
\begin{abstract}
A thermodynamically consistent density dependent quark mass model
has been used to study the velocity of sound in a hot quark-gluon plasma.
The results are compared with the recent lattice data.
\end{abstract}
\pacs{25.75.Nq;12.38.Mh;12.39.Ki} 

\maketitle
{\bf Key words:} Equation of state, Quark Gluon Plasma, Sound velocity 
\vskip 0.1in
{\bf To appear in Mod. Phys. Lett. A}
\vskip 0.1in

The study of the Equation of State (EOS) of strongly interacting matter 
is of great importance. It can play a vital role for a deeper 
understanding of many unresolved areas of QCD physics, such as the study of 
quark-gluon plasma (QGP), quark confinement, phase structure
of QCD at high temperature and finite chemical potential, to name a few.

The study of QGP has some formidable problems. Due to the convergence
problem in the low energy domain, the widely used perturbative QCD 
calculation fails badly to give any insight. The most accepted 
method in this area is the lattice QCD calculation, though the method 
works better with light dynamical quarks.
Moreover, there are difficulties with finite baryon chemical potentials.
On the other hand, various phenomenological models of thermodynamical and 
hydrodynamical types \cite{1,2,3,3a,4,5,6,7,7a,7b,8,9,10} are used as they are easier 
to handle compared to the
lattice. In all these models, a quasiparticle description 
is used with a background mean field. Here it is worth noting that a 
quasiparticle description is expected to be applicable, as long as 
the spectral functions for quarks and gluons
resemble qualitatively the asymptotic forms found from HTL (Hard Thermal
Loop) perturbative calculations \cite{11}.

The calculations in the quasiparticle description usually start with a
Hamiltonian. It was pointed out by Bir\(\grave{o} \) et al. 
\cite{12} that in a Hamiltonian approach, due to the dependence
of the Hamiltonian on the thermodynamical characteristics of the 
surrounding matter, there may be some thermodynamical inconsistency 
within the method which has to be rectified. To get rid of 
these inconsistencies, one has to incorporate certain  
conditions in the quasiparticle model.  

In all the studies of the EOS, the sound velocity (\( u \)), which relates the
pressure and the energy as \( P=u^2 E \), plays an important role. Specifically,
the QGP is expected to approach the canonical value of \( u^2 = 1/3 \), at
high temperatures and/or densities. An understanding of the quantitative 
behaviour of the approach to asymptotic freedom is very important, as it is
related to the problem of quark confinement.

Another important quantity related to the study of QGP is the quark number 
susceptibility (QNS), which, being related to charge fluctuation, is of 
direct experimental relevance. The importance of QNS lies in the fact that
it can be used as an independent check on the theoretical models which try to
explain the lattice results \cite{13}.

There are numerous results of quark number susceptibility available from 
lattice calculations \cite{13,14,15}. 
Also, There have been recent reports of lattice 
studies where the velocity of sound in a pure gluon plasma at finite 
temperature has been calculated \cite{16}. Our 
motivation here is to understand what kind of quasiparticle
picture can reproduce these results.

In the present work, we have studied the behaviour of quark number 
susceptibility
and the speed of sound as the quark-gluon plasma moves from the non-perturbative
 to the perturbative region of QCD. The density dependent quark mass (DDQM) 
model 
\cite{17,18}, along with the condition for thermodynamical consistency, has been 
used for the present study.
\par
The DDQM model aims to explain confinement from a phenomenological
point of view. The MIT bag model \cite{1} is so far the simplest option
 available to
explain the confinement phenomenon, though it has some serious shortcomings. The 
DDQM model is a simple but effective alternate approach to confinement. The 
main idea behind the DDQM model is to make the mass of the ``free" quarks 
infinitely
heavy (confinement) while at small distances, effective quark mass should be
small or zero \cite{19}.
This behaviour is similar to what is known as ``Archimedes Effect" \cite{20,21}.

In the DDQM model, confinement is achieved through a parametrization of the
effective quark mass as a function of density. In the low density limit, 
all the interactions are taken care of by the mass, so that it becomes 
infinitely
large to prevent the occurrence of a single quark in the asymptotic state 
\cite{17}.
The thermodynamical quantities related to the quark sector
at finite temperature are calculated from the postulated mass relation and a 
self-consistent number density relation. 
At finite temperature the quark mass is defined as \cite{18},
\begin{eqnarray}
m(T,n_q) &=& \frac{B}{ \Sigma_i (n_i^+ + n_i^-)},
\end{eqnarray}
where $n_i^+ (n_i^-)$ is the number density of the i-th flavour quark (antiquark) and 
$n_q$
is the net quark number density.
These number densities are to be calculated self-consistently from the given relation:
\begin{eqnarray}
n_i^ \pm &=& \frac{3}{\pi^2} \int_0^{\infty}  dp  p^2 \{ \exp[T^{-1} (\epsilon_i \mp \mu_i)] 
+1 \}^{-1},
\end{eqnarray}
with,
\begin{eqnarray}
\epsilon_i (p,n_q,T) =[p^2+m^2 (T,n_q) + \Delta_i (p)]^{1/2}
\end{eqnarray}
where $\Delta_i (p)$ takes care of the divergent term present in the mass at
finite temperature. In the zero density limit, $\Delta_i (0)=0$ as the mass 
takes care
of all interactions. The expression for $\Delta_i (p)$ is given by \cite{18}:
\begin{eqnarray}
d\Delta_i(p_{i}) \equiv \left\{ \frac{6B \pi^2 c_i}{{p_{i}}^4}  \times
\frac{[{p_{i}}^2 -(B\pi^2 c_i/{p_{i}}^3 )^2]}{[{p_{i}}^2
+(B\pi^2 c_i/{p_{i}}^3 )^2]^{1/2}} +\frac{6B^2 \pi^4 {c_i}^2}{{p_{i}}^7} 
\right\} dp_{i}
\end{eqnarray}
with $c_i \equiv n_i/n_q$.
The energy density and the pressure of the quark system are then readily 
calculated from
\begin{eqnarray}
{\cal E}_q = \displaystyle \sum_{i=u,d} \frac{3}{\pi^2} \int_0^{\infty} 
dp [p^2 \epsilon_i(
p,n_q,T) ( \{ \exp[T^{-1}(\epsilon_i -\mu_i)] +1\}^{-1} + 
\{ \exp[T^{-1} (\epsilon_i +\mu_i)] +1\}^{-1} )] 
\end{eqnarray}
\begin{eqnarray}
P_q = \frac{1}{3} \displaystyle \sum_{i=u,d} \frac{3}{\pi^2} \int_0^{\infty} p
 \frac{\partial \epsilon}
{\partial p} {d^3}p ( \{ \exp[T^{-1}(\epsilon_i -\mu_i)] +1\}^{-1} + 
\{ \exp[T^{-1} (\epsilon_i +\mu_i)] +1\}^{-1} )
\end{eqnarray}
In the gluonic sector, energy density and pressure have the form \cite{18}:
\begin{eqnarray}
{\cal E}_g = \frac{8}{15} \pi^2 T^4 (1- \frac{15 \alpha_c}{4 \pi}), \\ \nonumber
P_g = \frac{8}{45} \pi^2 T^4 (1- \frac{15 \alpha_c}{4 \pi})
\end{eqnarray}
Here, $\alpha_c$ is the effective gluon-gluon coupling. The form of $\alpha_c$
is given by \cite{18}:
\begin{eqnarray}
\alpha_c = \frac{54B\pi^3}{(<Q^2>_{n_q,T})^2}
\end{eqnarray}
where,
\begin{eqnarray}
<Q^2>_{n_q,T}= \frac{4}{3} \frac{\displaystyle \sum_i (n_i^+ + n_i^-)
<Q^2>_i + n_g <Q^2>_g}{\displaystyle \sum_i (n_i^+ + n_i^-) + n_g}
\end{eqnarray}
$<Q^2>_i$ and $<Q^2>_g$ are the thermal average of the three momentum squared 
of quarks of flavour i and gluons. $<Q^2>_g$ corresponds to non-interacting
gas of gluons, $n_g$ being the gluon number density \cite{18,22}.
\par
So the total energy density and pressure
of the quark-gluon plasma is,
\begin{eqnarray}
{\cal E}_{QGP} = {\cal E}_q + {\cal E}_g,
P_{QGP} = P_q + P_g,
\end{eqnarray}
In the regions of low temperature and density, the energy density and pressure 
of the gluonic sector becomes negative which can be interpreted as the signal 
of the formation of gluonic condensates. As long as the gluons remain in the condensed 
phase they do not contribute explicitly to the thermodynamic quantities 
(they are not active degrees of freedom of the system).
So,
${\cal E}_g$ = ${\cal P}_g$ =0  for $\alpha_c \geq \frac{4 \pi}{15}$ \cite{18}.
\par
The DDQM model has been widely used in the literature for the study of dense
quark matter \cite{23,24,25,26,27,28}, the thermodynamical 
inconsistency notwithstanding \cite{12}.
Let us now modify the above EOS incorporating the conditions
of thermodynamical consistency. The prescription for ensuring thermodynamical
consistency is to introduce a "background" field \( \Phi \) which should satisfy
the following conditions \cite{12};
\begin{eqnarray}
\frac{\partial \Phi}{\partial T} + \displaystyle \sum_{j=u,d}d_j \int \frac{d^3k}{(2 \pi)^3}
\frac{\partial \epsilon_{kj} }{\partial T} \nu_{kj} &=& 0, \\
\frac{\partial \Phi}{\partial n_i} + \displaystyle \sum_{j=u,d} d_j \int \frac{d^3k}{(2 \pi)^3}
\frac{\partial \epsilon_{kj} }{\partial n_i} \nu_{kj} &=& 0.
\end{eqnarray}
where $d_j$ is the degeneracy factor and $\nu_{kj}$ is the average quasiparticle 
occupation number of j-th 
flavour quark with momentum k:
\begin{eqnarray}
\nu_{kj}=( \{ \exp[T^{-1}(\epsilon_j -\mu_j)] +1\}^{-1} +
 \{ \exp[T^{-1} (\epsilon_j +\mu_j)] +1\}^{-1} )]
\end{eqnarray}
In the present case, the above two conditions
become,
\begin{eqnarray}
\frac{\partial \Phi}{\partial T} + \displaystyle \sum_{j=u,d} n^{(s)}_j \frac{\partial m_j }
{\partial T}  &=& 0, \\
\frac{\partial \Phi}{\partial n_i} + \displaystyle \sum_{j=u,d} n^{(s)}_j \frac{\partial m_j }
{\partial n_i}  &=& 0.
\end{eqnarray}
where $n^{(s)}_j$ is the scalar quasiparticle density of j-th flavour quark,
\begin{eqnarray}
n^{(s)}_j &=& d_j \int \frac{d^3 k}{(2 \pi)^3} \nu_{kj} \frac{m_j}{\epsilon_j}
\end{eqnarray}
In the present paper we have considered a two flavour QGP system. Since \( u \) and
and \( d \) quarks have equal masses, our system, in essence, consists of
one type of quasiparticles only (except for the degeneracy factor 2 for two flavours).
So, the quasiparticle quark mass depends on temperature and 
quasiparticle density through the scalar density only and the expression
for the background field $\Phi$ reduces to
\begin{eqnarray}
\Phi = - \int \displaystyle \sum_{j=u,d} n^{(s)}_j dn_j
\end{eqnarray}
Carrying out the integration and identifying the integration constant with
$\Phi=0$ at T=0, $n_B=0$, background field $\Phi$ can be obtained.
At zero temperature, $\Phi$, naturally, depends on the baryon density 
($n_B$) alone. 
With this background field, we define thermodynamically consistent energy density
and pressure as,
\begin{eqnarray}
{\cal E}_{QGP}= {\cal E}_q +{\cal E}_g + \Phi (n_B,T) \\
P_{QGP}= P_q +P_g - \Phi (n_B,T)
\end{eqnarray}
At zero temperature and density, the total energy density 
of the QGP becomes B, the energy of an empty bag. On the other hand, the pressure
of QGP becomes zero and we do not encounter the unphysical negative pressure
as found in most of the MIT bag like phenomenological models.
It is to be noted here that
there is no bag pressure in this model as such. Since the mass of the quark
becomes infinitely large in the density going to zero limit, 
a negative bag pressure would amount to double counting \cite{17}.
\par
We define the baryon (\( n_{i=0} \)) and isospin (\( n_{i=3} \))
densities and the corresponding susceptibilities as:
\begin{eqnarray}
n_i= \frac{T}{V}\frac{\partial ln \cal{Z}}{\partial \mu_i}, \ \ 
\chi_{ij}= \frac{T}{V}\frac{\partial^2 ln \cal{Z}}
{\partial \mu_i \partial \mu_j}.
\end{eqnarray}
where, \( \cal{Z} \) is the partition function and thermodynamic potential
\( \Omega \) is defined as \( \Omega = -P_{QGP} = -Tln \cal{Z} \). Also
\( \mu_{i=0} = \mu_u + \mu_d \) and \( \mu_{i=3} = \mu_u - \mu_d \). 
\par
In figure 1 and 2 we have plotted $\chi_3/\chi_{FFT}$ as a function of $T/T_c$.
Here $\chi_3$ is the iso-vector susceptibility and $\chi_{FFT}$ is the free 
field value; for details see \cite{29,30}.
It is to be noted here that the value of the critical temperature in figure 1 
and 2 are different. In figure 1 the $T_c$ is taken from the consideration
of the commonly held belief that the QCD critical temperature is between 
150-200 MeV; in our calculation it is taken to be 200 MeV and represented by
dash-dot curve. In figure 2 the critical temperature has been calculated from 
the DDQM (taking the zero pressure limit, beyond which quark matter system will
become unstable) and shown as continuous line in the graph.
The points are from lattice calculation \cite{13} for different
values of the valence quark mass. As seen from the
plot, our result is in qualitative agreement with the lattice result 
 in the region near the critical temperature.
But at higher temperature it approaches the ideal gas limit faster than what we
get from the lattice calculation. Here it is worth mentioning that in our 
calculation current quark mass is zero and the dynamic 
quark mass varies with temperature.

The variation of \( \Phi \) with temperature at three different baryon densities
for B$^{1/4}$=145 MeV is shown in figure 3. The value of \( \Phi \)
is maximum for \( n_B = 0 \) and it decreases for increasing density. 
This behaviour of of \( \Phi \) could have been anticipated, as the system is 
expected to be closer to the perturbative domain at higher densities.
Near the transition temperature
\( T_c \), \( \Phi \) shows a rapid variation and then saturates at high
temperature. In fact, around the saturation region the effect of \( \Phi \)
on thermodynamic quantities becomes negligibly small. 
In figure 4, we have plotted  ($\cal{E}$-3p)/T$^4 $ with temperature. 
The two curves
are for the cases with and without the background field \( \Phi \); B$^{1/4}$
 and \( n_B \) being fixed at 145 MeV and 3n$_0$ (where n$_0$ = 0.17
\( fm^{-3} \) is the nuclear matter saturation density) respectively. 
The continuous and the broken lines correspond to the cases with and without 
the background field contribution, respectively. Inclusion of the background 
field into the model increases the critical temperature by $29\%$  
in the DDQM. 
The non-perturbative effects are dominant near the transition
temperature.
The inclusion of thermodynamic consistency certainly enhances these
effects, as seen in figure 4. 
Similar behaviour is obtained for the other values of B as well.  
   
Our main motivation in this paper has been to study the velocity of sound, 
which is given by \( (\frac{\partial P}{\partial {\cal E}})_S \). 
But in the present paper we have plotted \( (\frac{\partial P}{\partial
{\cal E}})_T \), as it is easier to compute in the present model. 
Moreover, it was pointed out in  ref.\cite{18} that the difference 
between  \( (\frac{\partial P}{\partial{\cal E}})_T \) and  
\( (\frac{\partial P}{\partial{\cal E}})_S \) does not 
exceed 10\%. The figures 5, 6, 7 and 8 
show the variation \( (\frac{\partial P}{\partial {\cal E}})_T \) as a 
function of temperature for different densities at B$^{1/4}$=145 MeV. 
We have also presented the recent
results from lattice calculation \cite{16}. As mentioned earlier, the 
dash-dot curves in figures 6 and 8 are plotted with $T_c$ equal to 200
MeV. The continuous and the broken lines in figures 5 and 7 correspond
to the cases with and without the background field contribution, respectively, 
the $T_c$ being calculated from the DDQM.

As seen from all the plots, \( (\frac{\partial P}{\partial {\cal E}})_T \)
 is greatly modified by the
contribution from the background field around $T_c$; \( \Phi \) has a 
dominant effect near \( T_c \). Moreover, as
\( \Phi \) goes over to B at zero temperature and density, the change in
B affects the sound velocity much more than that due to the change in
density. It is to be noted here that the lattice results for sound velocity 
\cite{16}
presented here are for gluonic contribution only, whereas our results are for 
quark gluon plasma. 
The discontinuity in the plot is due to the built-in second order 
phase transition (gluon condensate) in the present EOS.

Here we would like to point out that in figures 7 and 8, we have compared 
our results for the speed of sound in QGP at non-zero \( n_B \) with the
lattice values at \( n_B = 0 \). These comparisons indicate that the
behaviour near \( T_c \) is governed mainly by the physics of confinement.
In our model, the presence of quarks (along with the gluons) has very
small effect on the in-medium behaviour of sound velocity.

To conclude, we have studied the velocity of sound in the QGP with a 
thermodynamically consistent DDQM model. This is a one parameter model and the 
dependence on the parameter is also seen to be very weak. This is the 
remarkable feature of this model. Incorporation of thermodynamic consistency
produces qualitatively and semi-quantitatively, the features of the lattice
data and also explains all the necessary features of the non-perturbative to
perturbative transition of the QCD physics. Our study shows that 
the incorporation of consistency condition is essential to understand the 
nonperturbative behaviour near transition temperature. The result obtained
for QNS falls within the range of lattice data. The result for the
sound velocity of quark gluon plasma for zero baryonic density 
at critical temperature 200 MeV is in qualitative agreement with the lattice 
result for pure gluon plasma. Our result of sound-velocity of QGP for the 
finite baryonic density may be considered as a prediction to be confirmed 
by the future lattice calculations.

\begin{figure}
\includegraphics{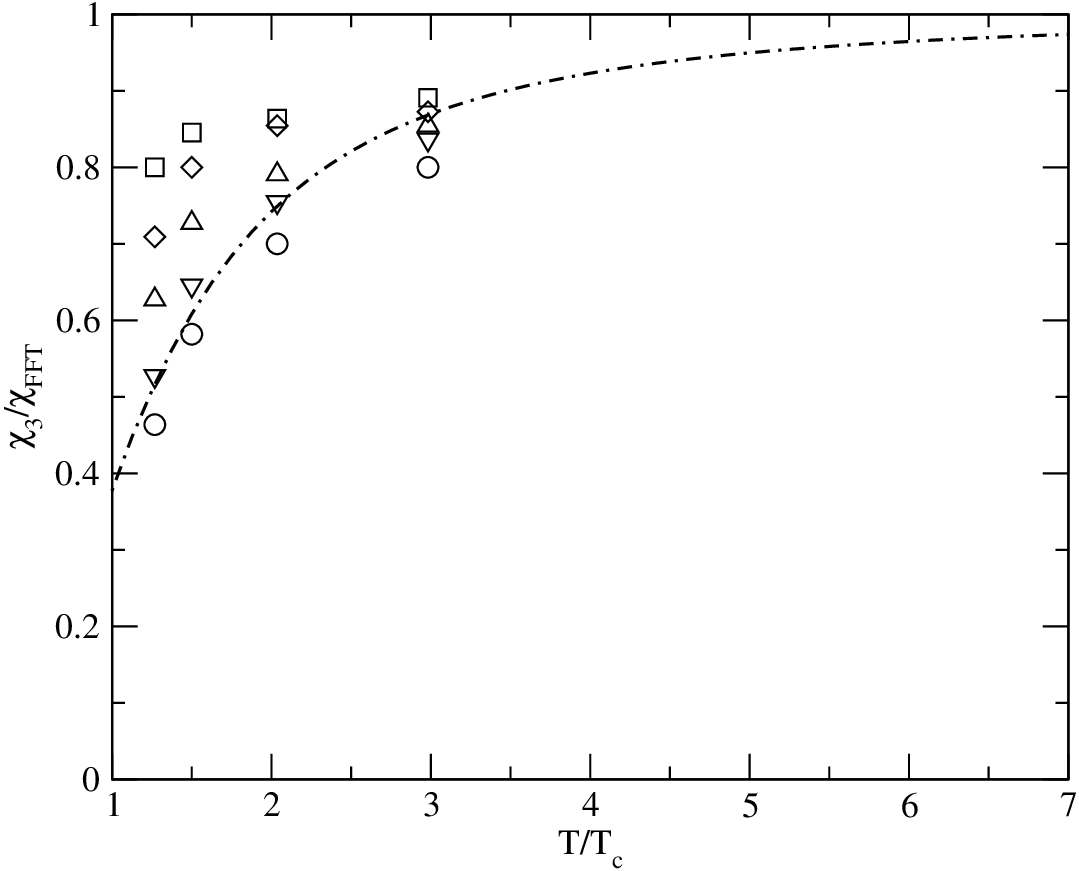}
\caption{\label{fig:chi-1} Variation of $\chi_3/\chi_{FFT}$
with temperature for B$^{1/4}$=145 MeV ,n$_B$=0 and $T_c= 200 MeV$. 
Dash-dot curve represents our model calculation. Different symbols like 
circle, down-triangle,
up-triangle,diamond and the square represent lattice data
corresponding to the valence quark mass 1.0, 0.75, 0.5, 0.3, 0.1, in units of 
$T_c$, respectively \cite{13}.} 
\end{figure}
\begin{figure}
\includegraphics{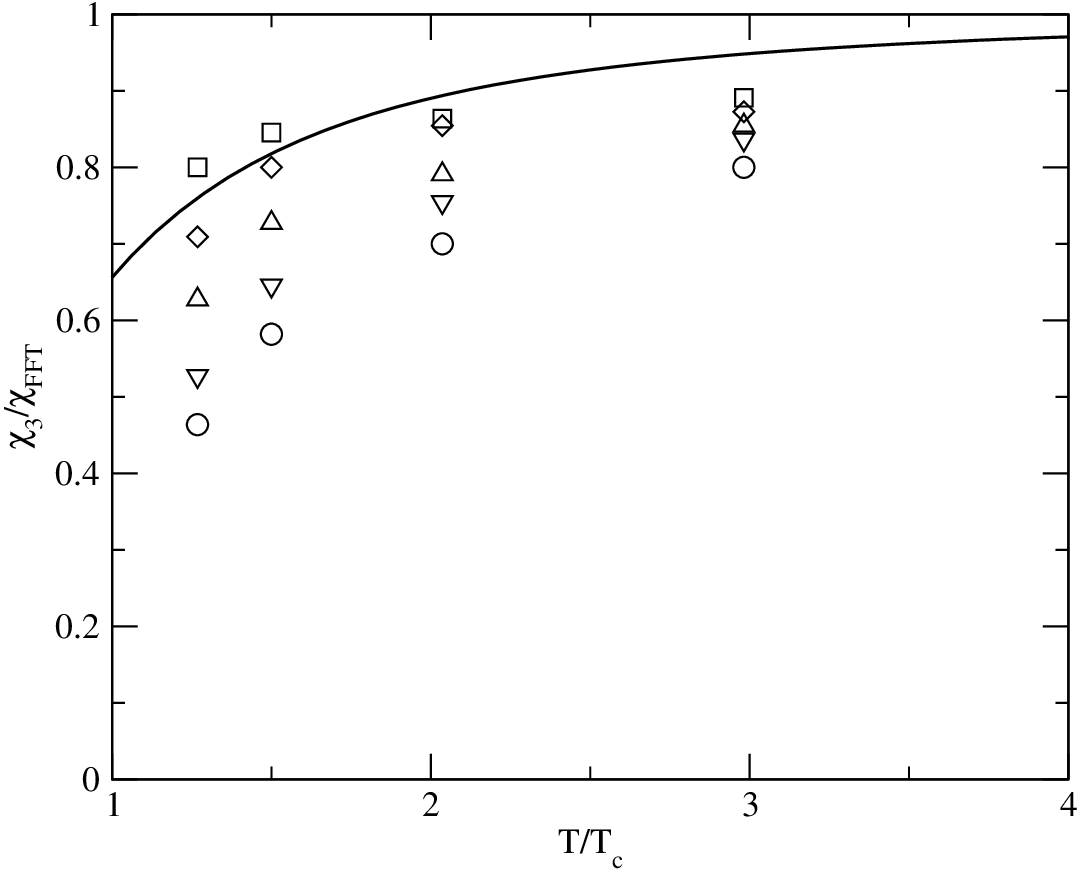}
\caption{\label{fig:chi-2} Variation of $\chi_3/\chi_{FFT}$
with temperature for B$^{1/4}$=145 MeV ,n$_B$=0 and T$_c$=330 MeV, calculated
from the model. Solid curve represents our model calculation. Different symbols
 like circle, down-triangle,
up-triangle, diamond and the square represent lattice data
corresponding to the valence quark mass 1.0, 0.75, 0.5, 0.3, 0.1, in units of 
$T_c$, respectively \cite{13}.} 
\end{figure}
\begin{figure}
\includegraphics{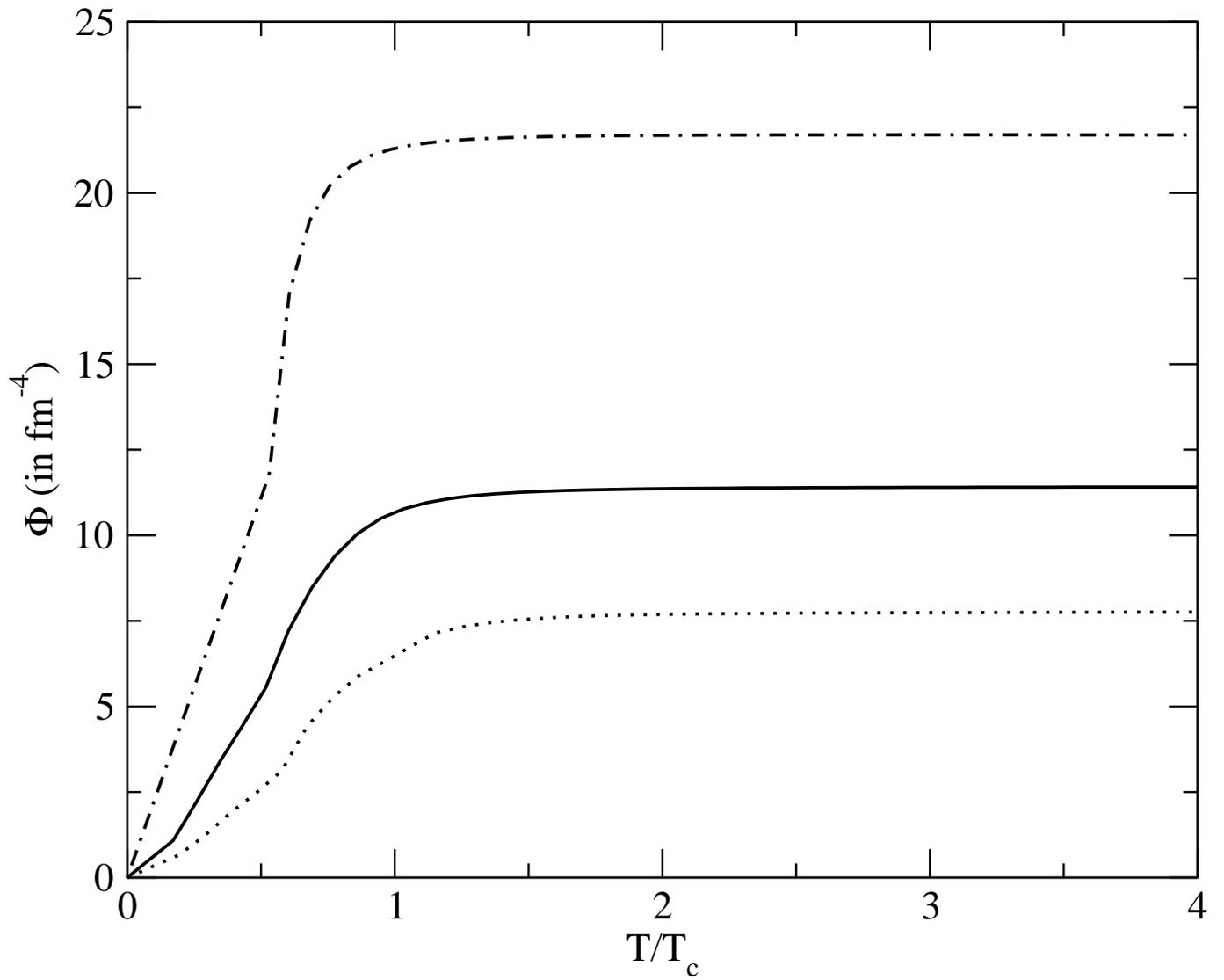}
\caption{\label{fig:phi} Variation of \( \Phi \) with temperature
for B$^{1/4}$=145 MeV. 
Dash-dot, solid and dotted curves are for densities n$_B$=0,  
n$_B$=3n$_0$ and n$_B$=4n$_0$
respectively, $n_0$ being the nuclear matter saturation density. Critical
temperature for n$_B$=0 is 330 MeV, for n$_B$=3n$_0$ it is 290 MeV and 
for n$_B=4n_0$ it is 260 MeV.} 
\end{figure}
\begin{figure}
\includegraphics{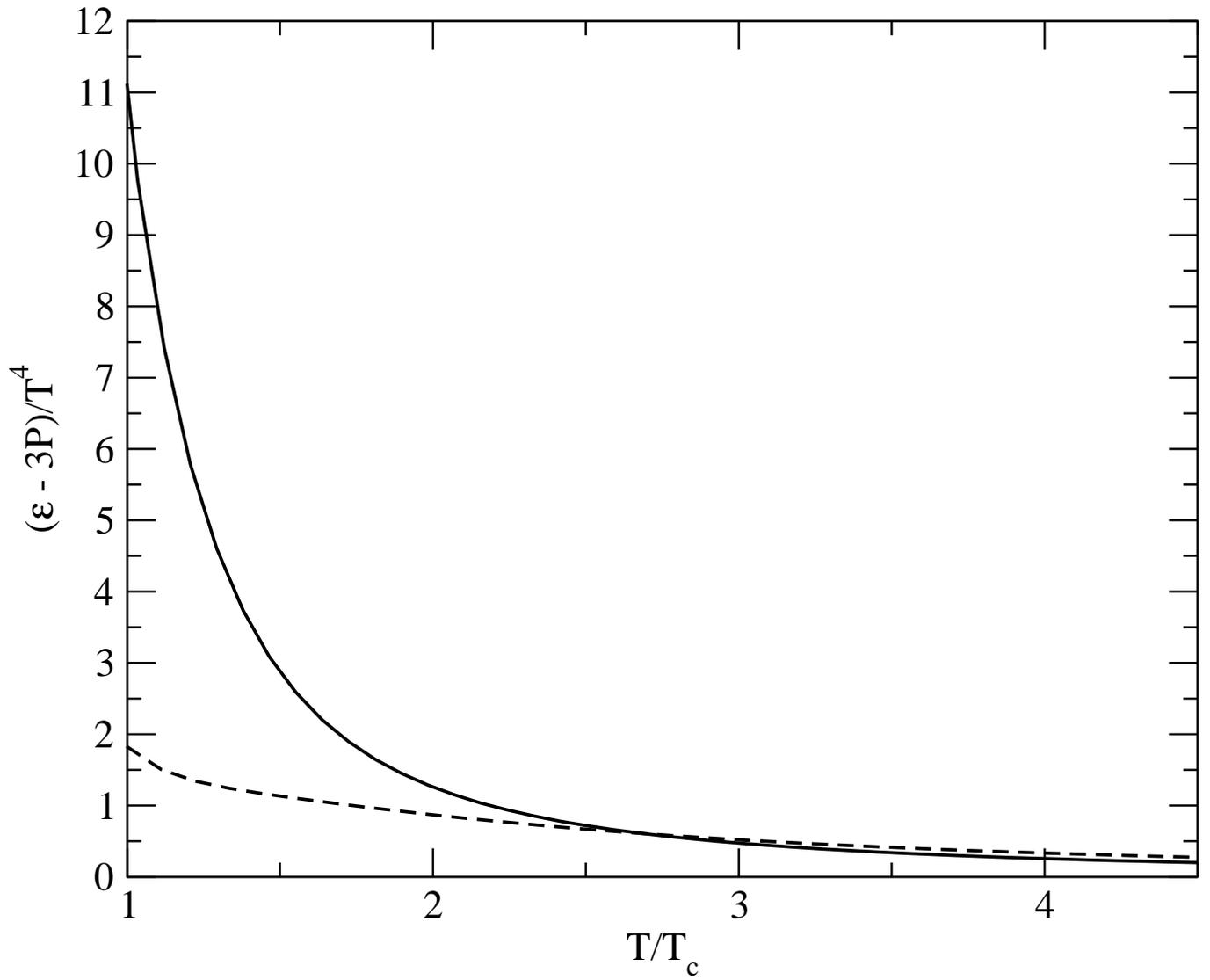}
\caption{\label{fig:e-p} Variation of ($\cal{E}$-3P)/T$^4 $ with temperature
for B$^{1/4}$=145 MeV, n$_B$=3n$_0$ and $T_c=290 MeV$. 
Solid and dash curves correspond to the cases with and without 
$\Phi$ contributions, respectively.}
\end{figure}
\begin{figure}
\includegraphics{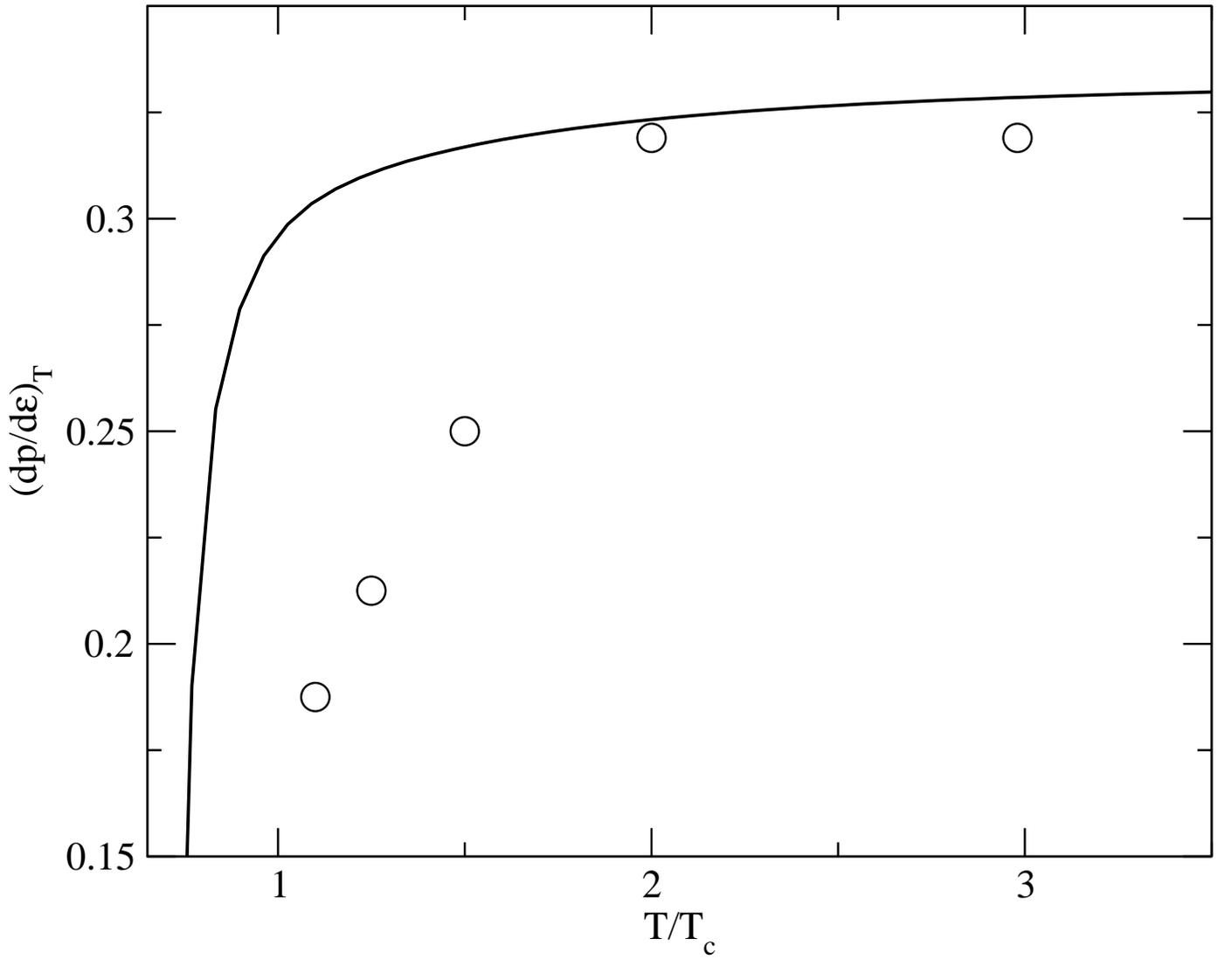}
\caption{\label{fig:s-v1} Variation of \( (\frac{\partial P}{\partial
{\cal E}})_T \) with temperature for B$^{1/4}$=145 MeV, n$_B$=0 and 
$T_c=330 MeV$
calculated from the model. Lattice data (circles) are taken from \cite{16}. 
Solid line is for our model.}
\end{figure}
\begin{figure}
\includegraphics{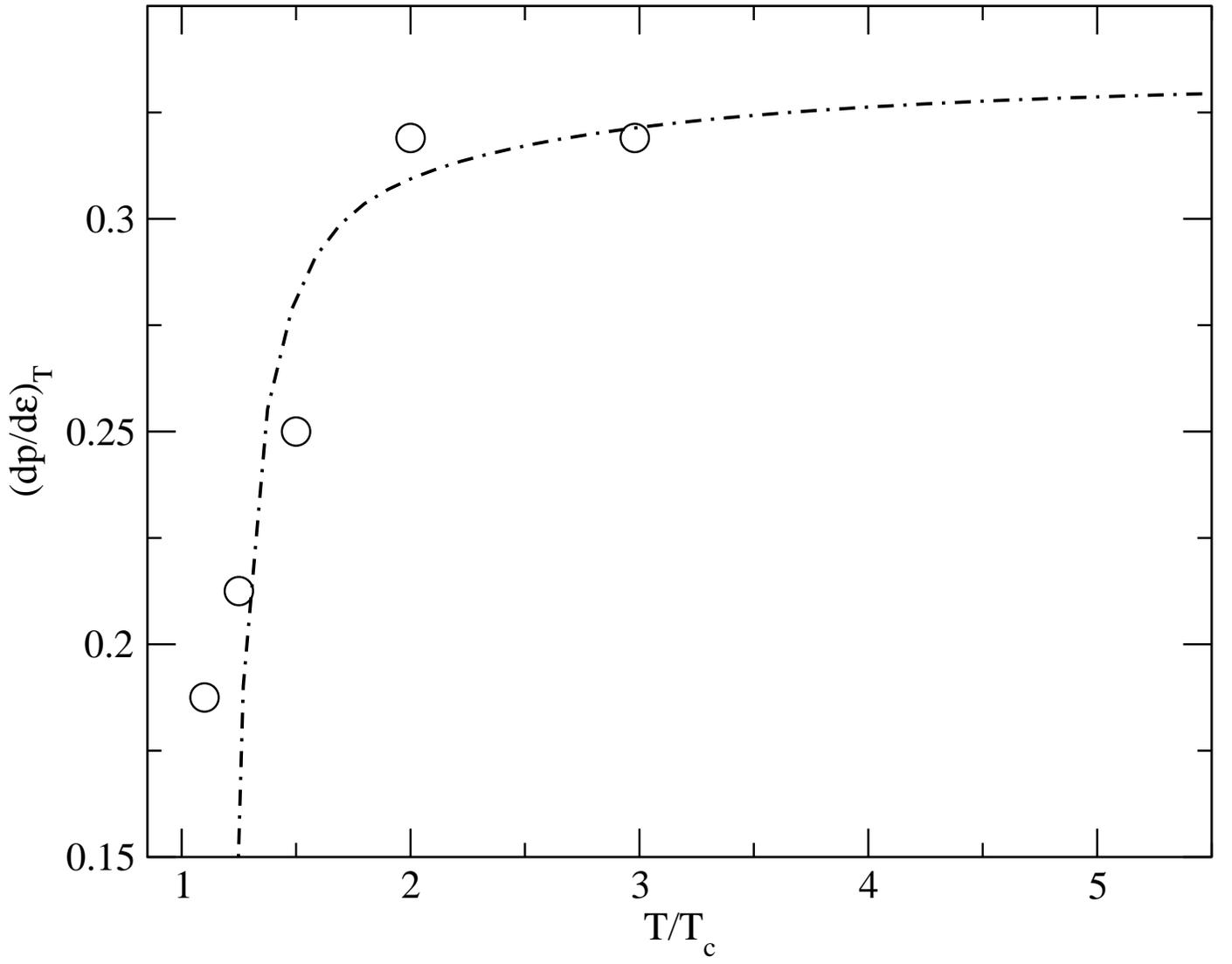}
\caption{\label{fig:s-v2} Variation of \( (\frac{\partial P}{\partial
{\cal E}})_T \) with temperature for B$^{1/4}$=145 MeV, n$_B$=0 and $T_c
=200 MeV$. Lattice data (circles) are taken from \cite{16}. Dash-dot line is for
 our model.}
\end{figure}
\begin{figure}
\includegraphics{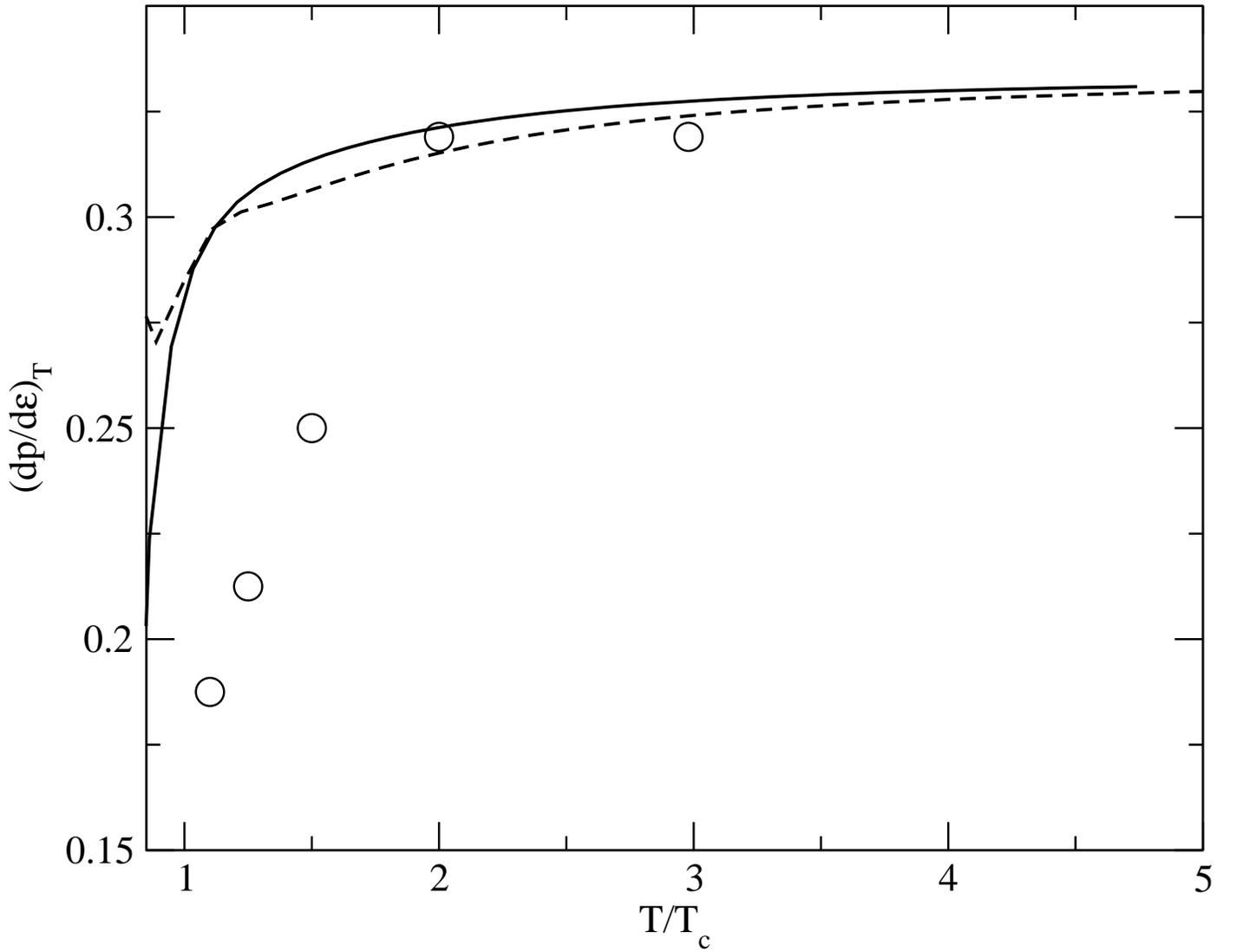}
\caption{\label{fig:s-v3} Variation of \( (\frac{\partial P}{\partial
{\cal E}})_T \) with temperature for B$^{1/4}$=145 MeV, n$_B$=3n$_0$ and $T_c
=290 MeV$, calculated from the model. Lattice data (circles) are taken from 
\cite{16}. Solid and the broken lines correspond to the cases with and without 
$\phi$ contributions, 
respectively. }
\end{figure}
\begin{figure}
\includegraphics{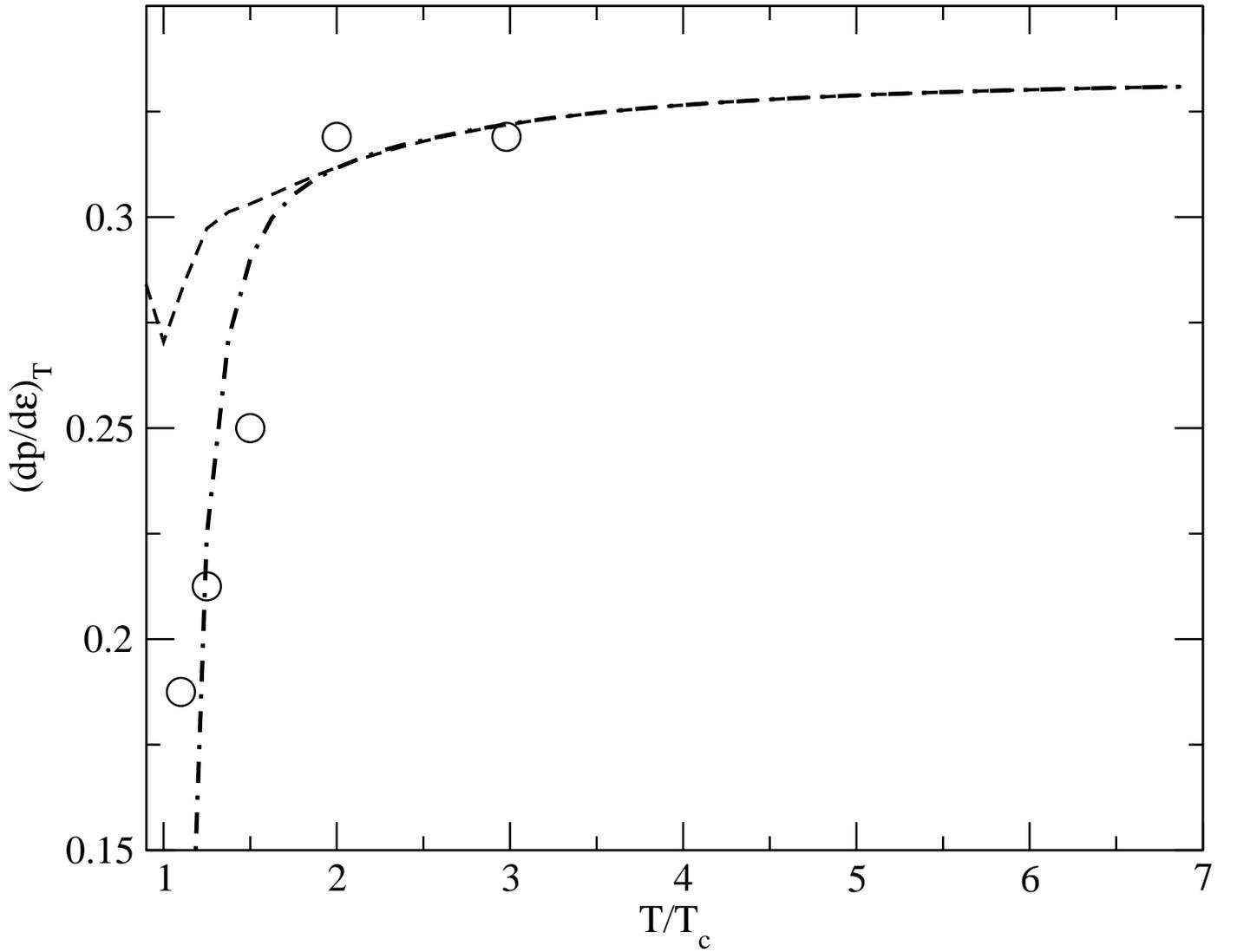}
\caption{\label{fig:s-v4} Variation of \( (\frac{\partial P}{\partial
{\cal E}})_T \) with temperature for B$^{1/4}$=145 MeV, n$_B$=3n$_0$ and $T_c
=200 MeV$. Lattice data (circles) are taken from \cite{16}. Dash-dot and the 
broken lines correspond to the cases with and without $\phi$ contributions, 
respectively.}
\end{figure}

\end{document}